\documentclass{llncs}
\usepackage{array,amsmath,amsfonts,amssymb,hyperref,alltt,makeidx,centernot,esvect,wasysym} 

\usepackage{color}
\usepackage{color,booktabs,fixltx2e,graphicx,etex,textcomp,caption}
\usepackage[linesnumbered,ruled]{algorithm2e}
\SetKwRepeat{Do}{do}{while}%

\usepackage{booktabs,fixltx2e,listings}
\usepackage{tikz-cd}
\usetikzlibrary{matrix,arrows,decorations.pathmorphing}
\usepackage{textcomp}
\usepackage{gnuplot-lua-tikz}
\usepackage{todonotes}
\usepackage{subcaption}

\newtheorem{thm}{Theorem}

\newcommand\mmit[1]{\ensuremath{\mathit{#1}}}

\newcommand{\La}[0]{\ensuremath{\langle}}
\newcommand{\Ra}[0]{\ensuremath{\rangle}}

\newcommand{\rankt} {\ensuremath{\mathit{rankt}}}

\newcommand{\comment}[1]{}
\newcommand{\ie}[0]{\emph{i.e.}, }
\newcommand{\eg}[0]{\emph{e.g.}, }

\newcommand\Comment[1]{}

\newcommand{\trans}[0]{\ensuremath{ \rightarrow }}

  \mathchardef \mhyphen="2D

\newcommand{\trs}[1]{\mbox{\ensuremath{\mathcal{M_{\text{#1}}}= \langle S_{#1}, \xrightarrow{#1}, L_{#1} \rangle}}}
\newcommand{\labf}{\ensuremath{\mathcal{L}}}
\newcommand{\M}[1]{\ensuremath{\mathcal{M}_{#1}}}
\newcommand{\Mp}[1]{\ensuremath{\mathcal{M}'}}
\makeatletter
\newcommand{\xrightarrowc}[2][]{\ext@arrow 0359\rightarrowfill@{#1}{\hspace{-4pt}#2}}
\makeatother

\makeatletter
\newdimen\arrow@ht
\setbox\@tempboxa\hbox{\(\xrightarrow{}\)}
\arrow@ht\ht\@tempboxa
\newdimen\plus@wd
\setbox\@tempboxa\hbox{\(\scriptstyle +\)}
\plus@wd\wd\@tempboxa
\def\righttransarrowfill@{\arrowfill@\relbar\relbar{\raisebox{0pt}[\arrow@ht][0pt]{\(\xrightarrow{}^+\hskip-\plus@wd\)}}}
\setbox\@tempboxa\hbox{\(\xrightarrow{}\)}
\arrow@ht\ht\@tempboxa
\newcommand{\tto}[2][]{\ext@arrow 0359\righttransarrowfill@{#1}{#2}\hskip\plus@wd}
\makeatother

\newcommand{\disjtrs}[3]{\ensuremath{\langle S_{#1} \uplus S_{#2},
\xrightarrow{#1} \uplus \xrightarrow{#2}, \labf{#3} \rangle }}

\let\orgautoref\autoref %

\renewcommand{\autoref}[1] {%
  \def\equationautorefname{Eq.}%
  \def\figureautorefname{Fig.}%
  \def\subfigureautorefname{Fig.}%
  \def\subfigureautorefname{Fig.}%
  \def\definitionautorefname{Definition}%
  \orgautoref{#1}%
}

\DeclareFontFamily{OT1}{bbm}{}
\DeclareFontShape{OT1}{bbm}{m}{n}{%
      <5> <6> <7> <8> <9> <10> <12> <17> gen * bbm%
      <10-12>bbm10%
      <12-17>bbm12%
      <17->bbm17%
      <-5>bbm5}{}

\DeclareSymbolFont{bm}{OT1}{bbm}{m}{n}
\DeclareMathSymbol{\N}{7}{bm}{'116}

\usepackage[printwatermark]{xwatermark}
\usepackage{xcolor}

\begin{document}
\title{An Efficient Runtime Validation Framework based on the Theory
  of Refinement} \author{Mitesh Jain \and Panagiotis Manolios}
\institute{Northeastern University}


\maketitle
\vspace{-.4cm}
\begin{abstract} 
  We introduce a new methodology based on refinement for testing the
  functional correctness of hardware and low-level software. Our
  methodology overcomes several major drawbacks of the de facto
  testing methodologies used in industry: (1) it is difficult to
  determine completeness of the properties and tests under
  consideration (2) defining oracles for tests is expensive and
  error-prone (3) properties are defined in terms of low-level
  designs. Our approach compiles a formal refinement conjecture into a
  runtime check that is performed during simulation. We describe our
  methodology, discuss algorithmic issues, and provide experimental
  validation using a 5-stage RISCV pipelined microprocessor and
  hypervisor.
\end{abstract} 

\vspace{-.4cm}
\section{Introduction}

Hardware and low-level software designs continue to increase in
complexity and as a result verification has become the dominant design
cost. According to a recent study by Foster about 70\% of designs
include embedded processors, SoC designs with over 100 IP blocks are
not uncommon, the mean number of verification engineers is more than
the mean number of design engineers, and even design engineers spend
about half their time on verification~\cite{foster2015}.

In this paper, we introduce a new approach to dynamic verification,
based on refinement. In the last decade, refinement-based methodology
has been successfully used to statically verify correctness of several
practical systems like pipelined microprocessors, operating systems
microkernels and distributed
systems~\cite{klein2010refinement,hawblitzel2015ironfleet}. Refinement
shows that the behaviors of a concrete system, say a pipelined
machine, are suitably related to the behaviors of an abstract machine,
say an instruction set architecture, that serves as the
specification~\cite{manolios2006framework}.
The idea behind our method is simple. We compile a refinement
conjecture into a runtime check that is performed during
simulation. This allows us to check for functional correctness during
testing using only this one check. To our knowledge, we are the first
to propose and study the application of a theory of refinement for
dynamic verification. Our approach addresses several challenges facing
industry~\cite{foster2009applied,armoni2002forspec,andersen2005leveraging,chatterjee2005streamline}.
First, we target functional correctness. According to Foster, 50\% of
flaws resulting in respins are due to logic or functional correctness
flaws.  Second, it is difficult to determine if the set of properties
and tests under consideration is complete: the Foster study that shows
that over 40\% of of functional flaws are due to incomplete or
incorrect specifications. Third, defining oracles for tests is
expensive and error-prone: the Foster study shows that verification
engineers spend 24\% of their time creating tests and running
simulations. Finally, properties and tests are defined in terms of
low-level designs, so modifications during the design cycle lead to,
possibly significant, changes to the properties being tested. The
Foster study shows that over 40\% of the functional flaws are due to
change in the specification. This is undesirable. Furthermore, the
effort to maintain properties for an industrial design can often be
very large~\cite{turumella2008assertion} and cannot be under
estimated.

We briefly review refinement in Section~\ref{sec:refinement}, followed
by the introduction of a running example in
Section~\ref{sec:example}. We present our refinement-based testing
method in Section~\ref{sec:testing}, which includes algorithms, a
theorem of correctness, some optimizations, and a discussion of the
overhead incurred by our method.  Experimental validation using two
case studies on pipeline machine verification and a case study on the
verification of a hypervisor is the topic of
Section~\ref{sec:evaluation}. After a discussion and related work, in
Section~\ref{sec:discussion}, we end with conclusions and future work
in Section~\ref{sec:conc}.

\vspace{-.3cm}
\section{Refinement}
\label{sec:refinement}

Refinement allows us to show that a low-level \emph{concrete} system
correctly implements a high-level \emph{abstract} system by showing
that every behavior of the concrete system is allowed by the abstract
system.  To bridge the gap between the concrete and abstract systems,
we use a \emph{refinement map}, a function that maps concrete states
to the abstract states they correspond to. For example, the commitment
refinement map~\cite{manolios2005computationally} takes as an input a
pipelined machine state and returns the ISA (Instruction Set
Architecture) state obtained by invalidating all partially executed
instructions and projecting out the programmer-observable components.

Many different notions of refinement exist
(\cite{manoliosPhd,manolios2015skipping}) and our approach is quite
general, but in this paper we will use WEB (Well-founded Equivalence
Bisimulation) refinement~\cite{manoliosPhd}, a notion of correctness
that is well studied in the context of hardware
verification~\cite{manolios2006refinement}.
In
this section, we provide a brief introduction to WEB refinement. We
start by defining transition systems (TS), a general,
language-agnostic setting, that we use to model systems.

\begin{definition}
  A TS \M{} is a three-tuple {$\La S, \trans, L \Ra$}, consisting of a
  set of states $S$, a transition relation $\trans$, and a labeling
  function $L$ whose domain is $S$.
\end{definition}

The labeling function is used to specify what is externally
observable in a state. Transition systems are both simple and
general, \eg $\trans$ is an arbitrary relation and there are no
cardinality restrictions on $S$. Thus transition systems can be
used to model infinite state machines and unbounded
nondeterminism.

We now define WEB-refinement, which is defined with respect to a
single transition system: the disjoint union ($\uplus$) of the
concrete and abstract TSs.

\begin{definition}
  Let \trs{C} be the transition system of a concrete system, \trs{A}
  be the transition system of an abstract system and
  $r: S_C \trans S_A$ be a refinement map. We say that \M{C} is a WEB
  refinement of \M{A} with respect to $r$, if there exists a binary
  relation $B$, such that $\La \forall s \in S_C:: sBr(s)\Ra$ and $B$
  is a WEB on TS \disjtrs{C}{A}{}, where $\labf{}(s) = L_A(r(s))$ for
  $s \in S_C$ and $\labf{}(s) = L_A(s)$ otherwise.
\end{definition}

The definition depends on the notion of a WEB (Well-founded
Equivalence Bisimulation)~\cite{manolios2000correctness}, which
given a binary relation $B$ checks that any pair of states
related by $B$ have identical behaviors up to finite stuttering.
\begin{definition}
  \label{def:webrefinementgeneral}
  $B \subseteq S \times S$ is a WEB on \trs{} iff:
  \vspace{-.3cm}
  \begin{enumerate}
  \item $B$ is an equivalence relation on $S$ and
  \item $\La \forall s,w \in S:sBw: L(s) = L(w)\Ra$ and
  \item There exist functions $rankl: S \times S \trans \mathbb{N}$,
    $rankt: S \times S \trans W$, such that $\La W, \prec \Ra$ is
    well-founded, and \\
    $\La \forall s,u,w \in S: sBw \wedge s \trans u:$
    \vspace{-.3cm}
    \begin{flalign*}
      & \textit{(a) }\La\exists v : w \trans v: uBv \Ra \; \vee&\\
      & \textit{(b) }(uBw \wedge rankt(u) \prec rankt(s)) \; \vee&\\
      & \textit{(c) }\La \exists v: w \trans v: sBv \wedge rankl(v,s) < rankl(w,s)\Ra\Ra&
    \end{flalign*}
  \end{enumerate}
\end{definition}

WEB refinement is equivalent to stuttering bisimulation
refinement~\cite{manolios2000correctness}, but requires only
local reasoning. This locality is the key that allows us to
design efficient algorithms for testing via refinement, because
it reduces the refinement problem, which is naturally expressed
in terms of infinite traces (that the behaviors of the concrete
system are allowed by the abstract system), to a problem
expressed in terms of states and their successors.

\vspace{-.3cm}
\section{Running Example}
\label{sec:example}
\vspace{-.3cm}
In this section, we describe our running example, a 3-stage pipeline
processor, MA, and the instruction set architecture, ISA, the abstract
specification for the MA machine. The running example will be used to
explain our refinement-based approach to testing. We say that the MA
\emph{implements} the ISA only if all observable behaviors of the MA
are behaviors of the ISA machine.

The stages of the MA pipelined processor are (1) fetch stage: the
machine fetches an instruction pointed to by the program counter (2)
load stage: the machine loads the source registers from the register
file or the data memory (3) execute stage: the machine executes the
instruction and updates the destination register in the register file
or data memory with the result.  The MA machine checks for data
dependencies between adjacent instructions in the pipeline; in case
the destination of the instruction in execute stage is equal to a
source register of the instruction in the load stage, the MA machine
stalls for one cycle. The ISA machine acts as the specification for
the MA machine. It just fetches the instruction pointed to by program
counter and executes it in a single step.
Notice that both the MA machine and the ISA machine are
deterministic.

\vspace{-.3cm}
\section{Testing via Refinement}  
\label{sec:testing}
In this section, we show how to test our running example using
WEB refinement.  The idea is simple: we simulate the design using
a test suite of programs. A program is run simultaneously on the
ISA and the MA machines and at each step, we check the refinement
conjecture.

Let us contrast our approach with prevailing methods used in industry,
which include the use of assertions and test
cases~\cite{foster2015}. To use these methods, a team of engineers is
needed to define a set of properties, a set of tests, and oracles that
determine when a test passes or fails. With our refinement-based
methodology, we only check one property, the WEB refinement property,
and the oracle is simply the high-level abstract model (ISA). Our
approach can therefore lead to significant savings of time and
effort. Our approach has the added advantage that refinement
completely characterizes functional correctness, whereas with
traditional methods, completeness is very difficult, if not
practically impossible, to achieve. We discuss this aspect in more
detail in Section~\ref{sec:discussion}.

We do not claim that our approach completely eliminates the need
for other testing methods. It does not, \eg low-level tests and
properties that check performance are needed. Also, we expect
that directed tests will be needed to achieve sufficient code
coverage. Nevertheless, we believe that the need for such tests
will be significantly reduced if refinement-based testing is
used. We provide evidence for this claim in
Section~\ref{sec:evaluation}.

We describe an algorithm for testing the WEB refinement property in
Algorithm~\ref{alg:general}. The algorithm is an online algorithm, \ie
we are checking refinement as the simulation is taking place. It is
possible to modify the algorithm to obtain an offline version, which
takes as input the trace of the concrete system and the abstract
system. The online version has several benefits over the offline
version, \eg it is simpler to describe, we do not need to store traces
and we can report errors as soon as they are discovered. An offline
algorithm may be easier to incorporate in an existing flow and may be
easier to parallelize. In section~\ref{sec:evaluation} We describe a
variation of the offline algorithm to check the functional correctness
of an open source pipeline processor.

We make one simplifying assumption in
Algorithm~\ref{alg:general}: no pair of abstract states are
labeled identically. This is a reasonable assumption, as usually
one models abstract states as tuples of state components with the
labeling function as the identity function. Such systems
trivially satisfy our assumption. With this assumption, we can
simply define the equivalence relation $B$ for checking WEB
refinement as the equivalence classes induced by the refinement
map.



Algorithm~\ref{alg:general} accepts as input an initial state $s$
of the concrete system; $n$, a bound on the number of simulation
steps to run; $r$, a refinement map; and $\mathit{rankt}$, a
function that maps states of the concrete system to a
well-founded domain (typically the natural numbers). We start by
initializing the abstract machine state $w$ to be $r(s)$ (\ie we
apply the refinement map to $s$), the $\mathit{error}$ flag to
$\mathit{false}$, the list $\mathit{partition}$ to $\La \Ra$ (the
empty list) and the two index variables $i$ and $j$ to $0$. Next
the algorithm loops as long as no error has been detected and
$n$, the bound on the number of simulation steps to take, is
positive. An inductive invariant at line~\ref{alg:gen:outerdo} is
$w=r(s)$.  The body of the loop starts by selecting $u$ to be a
successor state for $s$ (\emph{Select-concrete-next-state}). If
the concrete machine is nondeterministic, there can be many
successors of state $s$.  The correctness of the algorithm does
not depend on which state is selected, but as a practical matter,
it makes sense to select states randomly using an appropriate
distribution. We assume that we can replay the selections of $u$,
say because $u$ depends on a pseudo-random number generator, but
to simplify the presentation, this is not explicitly mentioned in
our algorithm.  Next the algorithm finds an abstract state $v$
that is a successor of $w$ and that matches $u$ (\ie $v=r(u)$),
if such a matching state exists, and sets $\mathit{match}$ to
$\mathit{true}$ (\emph{Match-abstract-next-state}). How this is
done depends on the abstract system. For example, if the abstract
system has bounded non-determinism, one simple strategy is to
iterate over the successors of $w$, stopping when a matching
state is found. If no such state exists, then $\mathit{match}$ is
set to $\mathit{false}$ and $v$ can be any abstract state. From
an algorithmic correctness perspective, it is crucial that
\emph{Match-abstract-next-state} always finds a matching state if
it exists. If such a $v$ exists, then the test in
line~\ref{alg:gen:ruv} holds and $u$ is matched in one step.  In
this case, we update \emph{partition}, a list of pairs of
indices, where the $i^{\mathit{th}}$ pair contains the indices of
states in the concrete and abstract traces corresponding to the
beginning of the $i^{\mathit{th}}$ partition.

\begin{thm}
If Algorithm~\ref{alg:general} reports an error, then the
concrete machine is not a WEB refinement of the abstract machine
with respect to the refinement map.
\end{thm}

Due to space limitations, the proof is omitted.  Notice that
given a refinement map $r$ and a rank function,
Algorithm~\ref{alg:general} only checks local conditions: the
relation between a concrete state $s$, its successor, an abstract
state $w$ and its successor. 

\begin{algorithm}[tb]
  \SetKwInOut{Input}{Input}
  \SetKwInOut{Output}{Output}
  \SetKwIF{If}{ElseIf}{Else}{if}{}{else if}{else}{}%
  \SetAlgoVlined
  \Input{$s$: concrete system state\\
    $n$: number of steps to run\\ 
    $r$: refinement map\\
    $\mathit{rankt}$: rank of concrete state\\ 
  } 
  \Output{Partition, Error Status}
  $w  \leftarrow r(s)$;
  $\mathit{error}  \leftarrow \mathit{false}$; 
  $\mathit{partition}  \leftarrow \La \Ra $;
  $i  \leftarrow 0$; $j  \leftarrow 0$\; 
  \Do{$n > 0 \wedge \neg \mathit{error}$}{ \label{alg:gen:outerdo}
    $u  \leftarrow \textit{Select-concrete-next-state}(s)$\; \label{alg:gen:select}
    \fbox{$\La match, v \Ra  \leftarrow \textit{Match-abstract-next-state}(w, u)$}\;
    \If{\fbox{match}}{ \label{alg:gen:ruv} 
      $partition  \leftarrow  partition :: \La i,j\Ra $\; 
      $i  \leftarrow i + 1$ ; $j  \leftarrow j + 1$\; 
      $s  \leftarrow u$ ; $w  \leftarrow v$\; 
    } 
    \ElseIf{\fbox{$r(u) = w \ \wedge \ \mathit{rankt}(u) \prec \mathit{rankt}(s)$}}{
      $i  \leftarrow i + 1$\; 
      $s  \leftarrow u$\; 
    } 
    \lElse{$\mathit{error}  \leftarrow \mathit{true}$}{
    }
    $n  \leftarrow n - 1$\;
  } 
  \Return $\La \mathit{partition}, \textit{error} \Ra$\; 
  \caption{Online WEB refinement Check}
  \label{alg:general}
\end{algorithm}
\vspace{-.5cm}

\subsection{Optimizations for Deterministic Machines}
 
Algorithm~\ref{alg:general} works even when the concrete and
abstract machines are nondeterministic. However, it is often the
case that the abstract and concrete machines are
deterministic. The algorithm can be simplified by exploiting
determinism. In Figure~\ref{fig:detwebrefinement} we show what
the WEB refinement conjecture can be simplified to when the
machines are deterministic and the abstract machine does not
stutter.  In the figure, $S_{C}$ is the set of states of the
concrete machine; $\textit{concrete-step}$ is the function
corresponding to one step of the concrete machine; $S_{A}$ is the
set of abstract states of the abstract machine; and
$\textit{abstract-step}$ is the step function of the abstract
machine. For deterministic machines, the
\textit{match-abstract-next-state} and
\textit{select-concrete-next-state} functions in
Algorithm~\ref{alg:general} can be replaced by functions
\textit{abstract-step} and \textit{concrete-step},
respectively. Furthermore, in case of deterministic machines, it
always suffices to use $\La \mathbb{N}, < \Ra$ as the
well-founded structure.

In our running example, the MA and ISA machine are both
deterministic and the ISA machine does not stutter; hence, we can
use the simplified algorithm. Next we define a refinement map
$r$, a function from MA states to ISA states that tells us what
is observable in a concrete state. Given an MA state, the
refinement map $r$ invalidates all partially completed
instructions in the pipeline, updates the program counter and
projects out the programmer observable components of the
state. If neither latch is valid (the pipeline is empty) then the
program counter stays the same. Otherwise, it points to the
oldest valid instruction in the pipeline (the instruction in the
second latch is older than the instruction in the first latch).
Furthermore, since no two states of the ISA machine are labeled
identically, we define $B$ to be the equivalence relation induced
by refinement map $r$. The choice of a refinement map plays a
crucial role in the efficiency of the algorithm. We discuss this
in detail in Section~\ref{sec:discussion}.
\vspace{-1cm}
\begin{figure}
  \hspace{2cm}
  \begin{minipage}{\textwidth}
    { \begin{flalign*}
     \La \forall s\in S_{C}::\ & u = \textit{concrete-step}(s)\; \wedge & \\
     & w = r(s) \; \wedge &\\
     & v = \textit{abstract-step}(w) \; \wedge v \centernot = r(u)&\\
     \Longrightarrow &\; w = r(u) \; \wedge \; \mathit{rankt}(u) <
     \mathit{rankt}(s)
   \end{flalign*}}
\end{minipage}
\vspace{-.5cm}
   \caption{\smaller{A simplified WEB refinement conjecture}}
   \label{fig:detwebrefinement}
 \end{figure}
\vspace{-.7cm}

Next we define the function $\mmit{rankt}$ in
Definition~\ref{def:webrefinementgeneral}; given an MA state $s$
\emph{rankt}($s$) is the number of MA steps required to commit an
instruction, \ie take an observable step, from $s$. If $s$ has a
valid instruction in the second latch, $\mmit{rankt} = 0$;
otherwise if there is a valid instruction in the latch for the
load stage, $\mmit{rankt} = 1$.
As we discuss later, the ranking functions are not essential for our
refinement-based testing method, but their use can help catch errors,
including performance errors. In case a rank function is not
available, the algorithm can still be used by removing all references
to $\mathit{rankt}$. The check relating $u$, and $w$ in
Algorithm~\ref{alg:general} is the safety component and the check
$ \mathit{rankt}(u) < \mathit{rankt}(s)$ is the liveness component.


\vspace{-.2cm}

\subsection{Overhead of checking WEB refinement}
We now discuss the overhead of checking for WEB refinement during 
simulation. The overhead costs of checking the WEB refinement
conjecture are indicated with a box in
Algorithm~\ref{alg:general}. The cost is primarily determined by the
following three factors: (1) computing the refinement map (2)
comparing the result of applying the refinement map to an abstract
state, and (3) computing the rank of a concrete state.

Refinement maps can be computationally expensive functions; hence, the
choice of refinement map plays a crucial role in the efficiency of the
algorithm. The refinement map described in Section~\ref{sec:testing}
is obtained by \emph{committing} an MA state and projecting out the
programmer visible components.  To efficiently compute the committed
state, we instrument the machine using \emph{history variables},
variable that record the history of programmer-observable components,
but do not affect the behavior of the processor. In the MA machine,
history variables record the values of the observable components of MA
states before they are updated in the pipeline. Notice that for the MA
machine, the program counter is the only observable component of the
concrete state that is updated by a partially executed instruction. We
evaluate the overhead cost of WEB-refinement checking during dynamic
validation in section~\ref{sec:evaluation} and refer the reader
to~\cite{manolios2005refinement,manolios2004automatic} for a detailed
comparison of computational efficiency of different refinement maps in
formal verification.

\vspace{-.3cm}
\section{Experimental Evaluation}
\label{sec:evaluation}
In this section we first evaluate the effectiveness of our method to
test the functional correctness of the MA machine, our running
example, and a small hypervisor~\cite{alkassar2010automated}.  We
evaluate the effectiveness of our method in detecting mutations and
the overhead costs of refinement checking during
simulation\footnote{\smaller{Experimental artifacts are available on
    request.}}. The MA machine and the hypervisor, and the
corresponding WEB refinement checkers based on
Algorithm~\ref{alg:general} are defined in ACL2s, the ACL2 Sedan. Next
we evaluate the ease of applying the refinement-based testing
methodology to an existing workflow. Towards this, we use the our
method to analyze the functional correctness of a 5-stage pipeline
processor based on RISC-V instruction set
architecture~\cite{riscv}. The processor model is described in Scala
and the refinement checker is implemented in Python.




\subsection{MA machine}
 \label{sec:pipeproc}
 In total we injected 25 errors in different components of the MA
 machine. These errors can roughly be classified into three classes:
 (1) \emph{Instruction classification}: mutations injected in
 classifying instructions as an arithmetic/logic unit (ALU)
 instruction, a load/store instruction, or a particular register is
 meaningful for the instruction; (2) \emph{Datapath}: mutations
 injected in computations in ALU and load/store unit (LSU). This class
 also includes mutations in connecting the ports of the modules; (3)
 \emph{Control logic}: mutations injected in detecting stall
 conditions and stalling the pipeline on data and structure hazards,
 error in branch recovery mechanism like invalidating instructions in
 the pipeline on the wrong path, etc.


 If we were to check functional correctness of the MA machine using
 the property-based approach, we would first specify a set of
 properties to ensure absence of each of the possible bugs. For
 example, we will have to specify a set of properties to ensure that
 the machine correctly handles pipeline hazards due to data
 dependencies between two instructions in the pipeline. For the MA
 machine, the property can be stated informally as follows: if the two
 pipeline latches have valid instructions, and if the older register
 has data dependency on the younger instruction, \ie the destination
 register of the older instruction is equal to any of the two source
 registers of the younger instruction, then a stall signal is asserted
 in the next cycle. In addition, we need another property that ensures
 that if the stall signal is asserted in a cycle, the program counter
 does not change in the next cycle. Similarly, we would need to
 specify a property for each possible scenario that we want to check
 during simulation. These properties are then compiled as runtime
 monitors that check for violations during simulations-based testing.

\begin{figure}[h]
  \begin{subfigure}{.5\textwidth}
    \hspace{-.5cm}
    \centering
    \scalebox{.7}{
      \begin{tabular}{|l|l|}
        \hline
        Errors Injected & Detect  \\ \hline
        \textbf{Instruction classification}               &                          \\ \hline
        1. sub not an alu-op         &  \checked    \\ \hline
        2. add not an alu-op         &  \checked    \\ \hline
        3. mul not an alu-op         &  \checked         \\ \hline
        4. load not a load-op        &  \checked    \\ \hline
        5. loadi not a load-op       &  \checked    \\ \hline
        6. store inst does not use rb  &  \checked                                   \\ \hline
        7. bez inst does not use rb  &  \texttimes                                   \\ \hline
        \textbf{Control: Stall Mechanism}                                         &         \\ \hline
        8. do not check if an inst is valid             &  \texttimes*     \\ \hline
        9. do not check if an inst has a valid dest reg     &  \texttimes*                           \\ \hline
        10. error in checking equality of dest and source reg ra   &  \checked    \\ \hline
        11. error in checking equality of dest and source reg rb   &  \checked           \\ \hline
        12. do not check if the op uses rbp  & \texttimes* \\ \hline
        \textbf{Control: Pipeline hazard detection}& \\ \hline                              
        13. error in invalidating younger inst on a taken branch             &  \checked    \\ \hline
        14. error in checking validity of a branch instruction       &  \texttimes    \\ \hline
        15. do not stall if the ra-val and rb-val are equal & \texttimes \\ \hline
        16. do not stall if ra-val  has a random value & \texttimes \\ \hline
        \textbf{Arithmetic-logic unit}                                  &                          \\ \hline
        17. error in addition          &  \checked     \\ \hline
        18. error in subtraction      &  \checked    \\ \hline
        19. error in multiplication    &  \checked         \\ \hline
        20. swapped ra and rb ports in a pipeline latch           &  \checked    \\ \hline
        21. swapped ra-val and rb-val ports in a pipeline latch  &  \checked    \\ \hline
        \textbf{Load store unit}                  &                    \\ \hline
        22. swap data and address ports for a store op    &  \checked    \\ \hline
        23. fuse data and address ports for a store op    & \checked     \\ \hline
        24. fuse data and address ports for a load op    &  \checked    \\ \hline
        25. error in address calculation for data memory access       &  \checked    \\ \hline
      \end{tabular}
    }
    \caption{\smaller{Mutations injected}}
    \label{tab:mamutations}
  \end{subfigure}%
  \begin{subfigure}{.5\textwidth}
    \vspace{2.6cm}
    \scalebox{.66}{
      \begin{tikzpicture}[gnuplot]
\gpsolidlines
\gpcolor{gp lt color border}
\gpsetlinetype{gp lt border}
\gpsetlinewidth{1.00}
\draw[gp path] (2.935,0.985)--(3.115,0.985);
\node[gp node right] at (2.751,0.985) { 0};
\draw[gp path] (2.935,2.218)--(3.115,2.218);
\node[gp node right] at (2.751,2.218) { 5};
\draw[gp path] (2.935,3.450)--(3.115,3.450);
\node[gp node right] at (2.751,3.450) { 10};
\draw[gp path] (2.935,4.683)--(3.115,4.683);
\node[gp node right] at (2.751,4.683) { 15};
\draw[gp path] (2.935,5.916)--(3.115,5.916);
\node[gp node right] at (2.751,5.916) { 20};
\draw[gp path] (2.935,7.148)--(3.115,7.148);
\node[gp node right] at (2.751,7.148) { 25};
\draw[gp path] (2.935,8.381)--(3.115,8.381);
\node[gp node right] at (2.751,8.381) { 30};
\draw[gp path] (2.935,0.985)--(2.935,1.165);
\node[gp node center] at (2.935,0.677) { 0};
\draw[gp path] (3.992,0.985)--(3.992,1.165);
\node[gp node center] at (3.992,0.677) { 2};
\draw[gp path] (5.048,0.985)--(5.048,1.165);
\node[gp node center] at (5.048,0.677) { 4};
\draw[gp path] (6.105,0.985)--(6.105,1.165);
\node[gp node center] at (6.105,0.677) { 6};
\draw[gp path] (7.162,0.985)--(7.162,1.165);
\node[gp node center] at (7.162,0.677) { 8};
\draw[gp path] (8.219,0.985)--(8.219,1.165);
\node[gp node center] at (8.219,0.677) { 10};
\draw[gp path] (9.275,0.985)--(9.275,1.165);
\node[gp node center] at (9.275,0.677) { 12};
\draw[gp path] (10.332,0.985)--(10.332,1.165);
\node[gp node center] at (10.332,0.677) { 14};
\draw[gp path] (2.935,8.381)--(2.935,0.985)--(10.332,0.985);
\node[gp node center,rotate=-270] at (1.861,4.683) {Simulation with WEB (seconds)};
\node[gp node center] at (6.633,0.215) {Simulation (seconds)};
\gpcolor{gp lt color 0}
\gpsetpointsize{6.00}
\gppoint{gp mark 1}{(3.595,1.589)}
\gppoint{gp mark 1}{(4.293,2.191)}
\gppoint{gp mark 1}{(4.969,2.795)}
\gppoint{gp mark 1}{(5.608,3.396)}
\gppoint{gp mark 1}{(5.614,3.707)}
\gppoint{gp mark 1}{(6.301,4.000)}
\gppoint{gp mark 1}{(6.618,4.338)}
\gppoint{gp mark 1}{(6.956,4.653)}
\gppoint{gp mark 1}{(7.315,4.954)}
\gppoint{gp mark 1}{(7.743,5.238)}
\gppoint{gp mark 1}{(7.896,5.600)}
\gppoint{gp mark 1}{(8.330,5.861)}
\gppoint{gp mark 1}{(9.122,6.202)}
\gppoint{gp mark 1}{(9.138,6.857)}
\gppoint{gp mark 1}{(9.434,6.966)}
\gppoint{gp mark 1}{(9.772,7.982)}
\gpcolor{gp lt color 1}
\gpsetlinetype{gp lt plot 1}
\draw[gp path] (3.595,1.525)--(3.658,1.584)--(3.720,1.642)--(3.783,1.701)--(3.845,1.760)%
  --(3.907,1.819)--(3.970,1.877)--(4.032,1.936)--(4.095,1.995)--(4.157,2.053)--(4.219,2.112)%
  --(4.282,2.171)--(4.344,2.230)--(4.407,2.288)--(4.469,2.347)--(4.531,2.406)--(4.594,2.464)%
  --(4.656,2.523)--(4.718,2.582)--(4.781,2.641)--(4.843,2.699)--(4.906,2.758)--(4.968,2.817)%
  --(5.030,2.875)--(5.093,2.934)--(5.155,2.993)--(5.218,3.052)--(5.280,3.110)--(5.342,3.169)%
  --(5.405,3.228)--(5.467,3.286)--(5.530,3.345)--(5.592,3.404)--(5.654,3.463)--(5.717,3.521)%
  --(5.779,3.580)--(5.841,3.639)--(5.904,3.698)--(5.966,3.756)--(6.029,3.815)--(6.091,3.874)%
  --(6.153,3.932)--(6.216,3.991)--(6.278,4.050)--(6.341,4.109)--(6.403,4.167)--(6.465,4.226)%
  --(6.528,4.285)--(6.590,4.343)--(6.652,4.402)--(6.715,4.461)--(6.777,4.520)--(6.840,4.578)%
  --(6.902,4.637)--(6.964,4.696)--(7.027,4.754)--(7.089,4.813)--(7.152,4.872)--(7.214,4.931)%
  --(7.276,4.989)--(7.339,5.048)--(7.401,5.107)--(7.464,5.165)--(7.526,5.224)--(7.588,5.283)%
  --(7.651,5.342)--(7.713,5.400)--(7.775,5.459)--(7.838,5.518)--(7.900,5.576)--(7.963,5.635)%
  --(8.025,5.694)--(8.087,5.753)--(8.150,5.811)--(8.212,5.870)--(8.275,5.929)--(8.337,5.987)%
  --(8.399,6.046)--(8.462,6.105)--(8.524,6.164)--(8.587,6.222)--(8.649,6.281)--(8.711,6.340)%
  --(8.774,6.398)--(8.836,6.457)--(8.898,6.516)--(8.961,6.575)--(9.023,6.633)--(9.086,6.692)%
  --(9.148,6.751)--(9.210,6.809)--(9.273,6.868)--(9.335,6.927)--(9.398,6.986)--(9.460,7.044)%
  --(9.522,7.103)--(9.585,7.162)--(9.647,7.220)--(9.710,7.279)--(9.772,7.338);
\gpcolor{gp lt color border}
\gpsetlinetype{gp lt border}
\draw[gp path] (2.935,8.381)--(2.935,0.985)--(10.332,0.985);
\gpdefrectangularnode{gp plot 1}{\pgfpoint{2.935cm}{0.985cm}}{\pgfpoint{10.332cm}{8.381cm}}
\end{tikzpicture}
    }
    \hspace{1.2cm}\caption{\smaller{Overhead cost of refinement checker}}
    \label{fig:runningtime}
\end{subfigure}
\caption{\smaller{Refinement checker for MA machine}}
\end{figure}
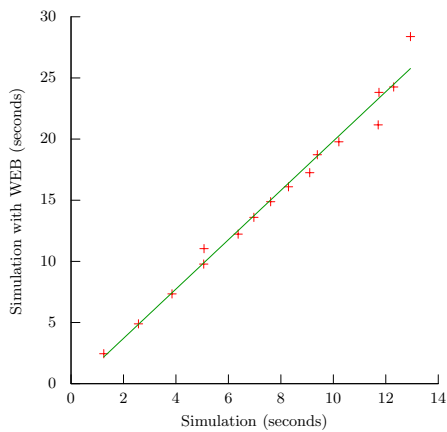

 In contrast, in our approach we specify and check the WEB refinement
 conjecture (Figure \ref{fig:detwebrefinement}) using a test program
 as input and Algorithm~\ref{alg:general}. Our test suite consisted of
 four generic programs: copy an array of memory from one location to
 another, perform multiplication by iterative addition, perform
 exponentiation by iterative multiplication, and a short sequence of
 10 random addition and subtraction instructions. Notice that our test
 programs are generic and are not crafted to find any particular
 error. In an industrial setting, we would expect to have a larger set
 of programs, which will only make it easier to find errors. Table
 \ref{tab:mamutations} lists the errors injected in the MA machine. If
 the bug was detected by refinement testing, we indicate this with a
 \checked \; mark in the second column of the table. In total 18 of
 the 25 injected errors were detected. Notice that three of the errors
 marked * in Table \ref{tab:mamutations} are not functional bugs and all
 of these errors falsely detect pipeline data hazards, resulting in
 stalling behavior that is not necessary. This is a loss of
 performance, but not a functional failure. Error in decoding that
 \emph{bez} (a conditional branch instruction) uses \emph{rb} as one
 of the source register (mutation 7) and error in checking the
 validity of a branch instruction while invalidating a pipeline latch
 (mutation 14) are functional bugs that with a more well-rounded set
 of programs are easy to detect.  We also added two
 difficult-to-detect mutations. One is a failure to stall the pipeline
 when the values of the two operands are equal (mutation 15) or when
 the value of an operand is an arbitrary value (mutation 16). The
 probability of stumbling on these bugs during testing is low (as is
 the probability of a designer introducing these bugs). Finding such
 errors requires the use of advanced counter-example generation
 techniques~\cite{chamarthi2011automated}.

We now analyze the overhead cost to check for WEB refinement in
simulation. In Figure~\ref{fig:runningtime}, we plot the running times
for simulating the MA machine ($x$-axis) vs the running times for
simulation with refinement testing enabled ($y$-axis). The points
shown indicate simulations ranging from 10,000 to 100,000 steps. The
slop of the fitting line is $\sim$2, which amounts to a slowdown of
approximately 100\%. Also notice that the points in the plot are on
either on the fitting line or near it, indicating that the overhead of
the checker does not increase as we increase the number of steps in a
simulation run.


\subsection{Simple Hypervisor}
\label{sec:hypervisor}
A hypervisor enables multiple operating systems (guests) to share
resources without interfering with one another. It achieves this by
virtualizing the host processor and the memory; a guest executing on a
virtualized system only exhibits behaviors that are admissible when
the guest is executing directly on the hardware. In this case study,
we implement a model of a simple
hypervisor~\cite{alkassar2010automated} and check its functional
correctness using WEB refinement in ACL2s.

A run of the virtualized system is defined as follows: The host
processor is time shared among multiple guests. At any given time,
only one guest is active (\textit{current guest}) on the host
processor. The hypervisor configures the host processor to execute in
user mode, irrespective of the mode of operation requested by the
current guest, and to use the shadow page-table of the current guest
in the host memory. The current guest directly executes instructions
in the instruction memory on the host processor until execution of an
instruction results in an exception. In case the exception is
``real'', \ie it is not a result of the guest executing in a
virtualized environment, the hypervisor passes the exception to the
guest. The exception is then handled by the exception handler
designated by the guest. Otherwise, the exception handled by the
hypervisor; the hypervisor saves the state of the host processor in
the processor control block (\emph{pcb}) for the current guest,
emulates the instruction, updates the \emph{pcb} and shadow page-table
(if required), and finally restores the state of the host processor
using the updated \emph{pcb}. The hypervisor hosts each guest memory
into disjoint regions of the host memory\footnote{For simplicity we
  assume that host memory is greater than sum of all guest memory plus
  memory required to store data-structures for the hypervisor.} and
controls the translation from a guest virtual address to physical
address.


A simple high-level abstract system where each guest executes on a
dedicated processor and memory, serves as a specification for the
virtualized system. Informally, the virtualized system is correct if
an observable behavior of a guest in the virtualized system is
identical to the observable behavior of the guest in the abstract
system. We encode the function correctness of the virtualized system
using WEB refinement. First we define the refinement map; it extracts
the state of the current guest from combination of the state of the
host processor and the current guest's \emph{pcb} in the
hypervisor. For all other guests, the corresponding pcb in the
hypervisor completely specifies the state of the guest in the abstract
system. The refinement map also extracts the guest memory partition
for each of the guest from the host memory. Next we define the
well-founded structure and the ranking functions. Notice that the
abstract system does not stutter. Hence, like in the pipeline
processor case study, we only need to define one ranking function,
\mmit{rankt} to account for stuttering in the concrete
system. Furthermore, since the concrete system is deterministic, it
suffices to use $\La \mathbb{N}, < \Ra$ as the well-founded structure.
We define the \mmit{rankt} as follows: if there is no pending
exception in a MA state, the rank is 1 else it is 0. Under these
conditions, we can use the simplified WEB refinement in
Figure~\ref{fig:detwebrefinement}.



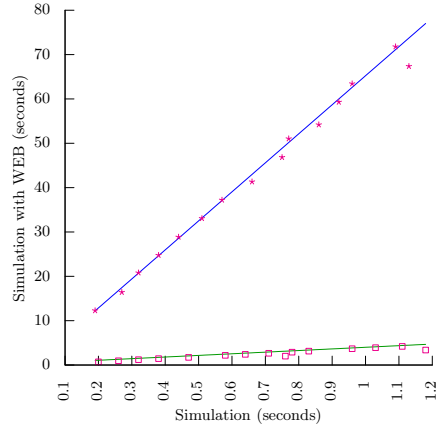
\begin{figure}
\vspace{-.3cm}
\begin{subfigure}{.5\textwidth}
  \centering \scalebox{.66}{
    \begin{tabular}{|l|c|}
      \hline
      Errors Injected & Detect\\
      \hline
      \textbf{Save/Restore} & \\ \hline
      1. wrong pc for inst responsible for exception & \checked\\ \hline
      2. wrong page-table origin & \checked\\ \hline
      3. check for host mode instead of guest mode & \checked\\ \hline
      \textbf{Emulation of Privileged Inst} & \\ \hline
      4. wrong pc used to fetch the inst to emulate & \checked\\ \hline
      5. do not increment pc on a move immediate value to PTO & \checked\\ \hline
      6. do not increment pc on a move immediate value to PTL & \checked\\ \hline
      \textbf{Sync Shadow Page Table} & \\ \hline
      7. use old pto to sync GPT and SPT & \checked\\ \hline
      8. use old ptl to sync GPT and SPT & \checked\\ \hline
      \textbf{Address calculation} & \\ \hline
      9. physical addr of guest pte in handling page-faults & \checked\\ \hline
      10. addr of page-table entry in host page-table & \checked\\ \hline
      11. addr of page-table entry in SPT & \checked\\ \hline
      \textbf{Hosting multiple guests}& \\ \hline
      12. Scheduler preempting a guest & \checked\\ \hline
      13. Guest memory separation & \checked\\ \hline
      \textbf{Others} & \\ \hline
      14. Fail to update the exception status after emulation& \checked \\ \hline
    \end{tabular}
  }
  \caption{\smaller{Mutations injected }}
  \label{tab:hyp-mutations}
\end{subfigure}%
\begin{subfigure}[]{.5\textwidth}
  \vspace{.1cm}
  \scalebox{.66}{
    \begin{tikzpicture}[gnuplot]
\path (0.000,0.000) rectangle (12.500,8.750);
\gpcolor{gp lt color border}
\gpsetlinetype{gp lt border}
\gpsetlinewidth{1.00}
\draw[gp path] (2.843,1.229)--(3.023,1.229);
\node[gp node right] at (2.659,1.229) {$0$};
\draw[gp path] (2.843,2.123)--(3.023,2.123);
\node[gp node right] at (2.659,2.123) {$10$};
\draw[gp path] (2.843,3.017)--(3.023,3.017);
\node[gp node right] at (2.659,3.017) {$20$};
\draw[gp path] (2.843,3.911)--(3.023,3.911);
\node[gp node right] at (2.659,3.911) {$30$};
\draw[gp path] (2.843,4.805)--(3.023,4.805);
\node[gp node right] at (2.659,4.805) {$40$};
\draw[gp path] (2.843,5.699)--(3.023,5.699);
\node[gp node right] at (2.659,5.699) {$50$};
\draw[gp path] (2.843,6.593)--(3.023,6.593);
\node[gp node right] at (2.659,6.593) {$60$};
\draw[gp path] (2.843,7.487)--(3.023,7.487);
\node[gp node right] at (2.659,7.487) {$70$};
\draw[gp path] (2.843,8.381)--(3.023,8.381);
\node[gp node right] at (2.659,8.381) {$80$};
\draw[gp path] (2.843,1.229)--(2.843,1.409);
\node[gp node right,rotate=-270] at (2.843,1.045) {$0.1$};
\draw[gp path] (3.515,1.229)--(3.515,1.409);
\node[gp node right,rotate=-270] at (3.515,1.045) {$0.2$};
\draw[gp path] (4.188,1.229)--(4.188,1.409);
\node[gp node right,rotate=-270] at (4.188,1.045) {$0.3$};
\draw[gp path] (4.860,1.229)--(4.860,1.409);
\node[gp node right,rotate=-270] at (4.860,1.045) {$0.4$};
\draw[gp path] (5.533,1.229)--(5.533,1.409);
\node[gp node right,rotate=-270] at (5.533,1.045) {$0.5$};
\draw[gp path] (6.205,1.229)--(6.205,1.409);
\node[gp node right,rotate=-270] at (6.205,1.045) {$0.6$};
\draw[gp path] (6.878,1.229)--(6.878,1.409);
\node[gp node right,rotate=-270] at (6.878,1.045) {$0.7$};
\draw[gp path] (7.550,1.229)--(7.550,1.409);
\node[gp node right,rotate=-270] at (7.550,1.045) {$0.8$};
\draw[gp path] (8.223,1.229)--(8.223,1.409);
\node[gp node right,rotate=-270] at (8.223,1.045) {$0.9$};
\draw[gp path] (8.895,1.229)--(8.895,1.409);
\node[gp node right,rotate=-270] at (8.895,1.045) {$1$};
\draw[gp path] (9.568,1.229)--(9.568,1.409);
\node[gp node right,rotate=-270] at (9.568,1.045) {$1.1$};
\draw[gp path] (10.240,1.229)--(10.240,1.409);
\node[gp node right,rotate=-270] at (10.240,1.045) {$1.2$};
\draw[gp path] (2.843,8.381)--(2.843,1.229)--(10.240,1.229);
\node[gp node center,rotate=-270] at (1.953,4.805) {Simulation with WEB (seconds)};
\node[gp node center] at (6.541,0.215) {Simulation (seconds)};
\gpcolor{gp lt color 1}
\draw[gp path] (3.448,1.322)--(3.515,1.325)--(3.583,1.328)--(3.650,1.331)--(3.717,1.335)%
  --(3.784,1.338)--(3.852,1.341)--(3.919,1.344)--(3.986,1.348)--(4.053,1.351)--(4.121,1.354)%
  --(4.188,1.358)--(4.255,1.361)--(4.322,1.364)--(4.390,1.367)--(4.457,1.371)--(4.524,1.374)%
  --(4.591,1.377)--(4.659,1.380)--(4.726,1.384)--(4.793,1.387)--(4.860,1.390)--(4.928,1.394)%
  --(4.995,1.397)--(5.062,1.400)--(5.129,1.403)--(5.197,1.407)--(5.264,1.410)--(5.331,1.413)%
  --(5.398,1.416)--(5.466,1.420)--(5.533,1.423)--(5.600,1.426)--(5.667,1.430)--(5.735,1.433)%
  --(5.802,1.436)--(5.869,1.439)--(5.936,1.443)--(6.004,1.446)--(6.071,1.449)--(6.138,1.452)%
  --(6.205,1.456)--(6.273,1.459)--(6.340,1.462)--(6.407,1.466)--(6.474,1.469)--(6.542,1.472)%
  --(6.609,1.475)--(6.676,1.479)--(6.743,1.482)--(6.810,1.485)--(6.878,1.488)--(6.945,1.492)%
  --(7.012,1.495)--(7.079,1.498)--(7.147,1.502)--(7.214,1.505)--(7.281,1.508)--(7.348,1.511)%
  --(7.416,1.515)--(7.483,1.518)--(7.550,1.521)--(7.617,1.524)--(7.685,1.528)--(7.752,1.531)%
  --(7.819,1.534)--(7.886,1.538)--(7.954,1.541)--(8.021,1.544)--(8.088,1.547)--(8.155,1.551)%
  --(8.223,1.554)--(8.290,1.557)--(8.357,1.560)--(8.424,1.564)--(8.492,1.567)--(8.559,1.570)%
  --(8.626,1.574)--(8.693,1.577)--(8.761,1.580)--(8.828,1.583)--(8.895,1.587)--(8.962,1.590)%
  --(9.030,1.593)--(9.097,1.596)--(9.164,1.600)--(9.231,1.603)--(9.299,1.606)--(9.366,1.610)%
  --(9.433,1.613)--(9.500,1.616)--(9.568,1.619)--(9.635,1.623)--(9.702,1.626)--(9.769,1.629)%
  --(9.837,1.632)--(9.904,1.636)--(9.971,1.639)--(10.038,1.642)--(10.106,1.645);
\gpcolor{gp lt color 2}
\draw[gp path] (3.448,2.321)--(3.515,2.380)--(3.583,2.438)--(3.650,2.497)--(3.717,2.555)%
  --(3.784,2.614)--(3.852,2.672)--(3.919,2.731)--(3.986,2.789)--(4.053,2.848)--(4.121,2.906)%
  --(4.188,2.965)--(4.255,3.023)--(4.322,3.082)--(4.390,3.140)--(4.457,3.199)--(4.524,3.257)%
  --(4.591,3.316)--(4.659,3.374)--(4.726,3.433)--(4.793,3.491)--(4.860,3.550)--(4.928,3.608)%
  --(4.995,3.667)--(5.062,3.725)--(5.129,3.784)--(5.197,3.842)--(5.264,3.901)--(5.331,3.959)%
  --(5.398,4.018)--(5.466,4.076)--(5.533,4.135)--(5.600,4.193)--(5.667,4.252)--(5.735,4.310)%
  --(5.802,4.369)--(5.869,4.428)--(5.936,4.486)--(6.004,4.545)--(6.071,4.603)--(6.138,4.662)%
  --(6.205,4.720)--(6.273,4.779)--(6.340,4.837)--(6.407,4.896)--(6.474,4.954)--(6.542,5.013)%
  --(6.609,5.071)--(6.676,5.130)--(6.743,5.188)--(6.810,5.247)--(6.878,5.305)--(6.945,5.364)%
  --(7.012,5.422)--(7.079,5.481)--(7.147,5.539)--(7.214,5.598)--(7.281,5.656)--(7.348,5.715)%
  --(7.416,5.773)--(7.483,5.832)--(7.550,5.890)--(7.617,5.949)--(7.685,6.007)--(7.752,6.066)%
  --(7.819,6.124)--(7.886,6.183)--(7.954,6.241)--(8.021,6.300)--(8.088,6.358)--(8.155,6.417)%
  --(8.223,6.475)--(8.290,6.534)--(8.357,6.592)--(8.424,6.651)--(8.492,6.709)--(8.559,6.768)%
  --(8.626,6.827)--(8.693,6.885)--(8.761,6.944)--(8.828,7.002)--(8.895,7.061)--(8.962,7.119)%
  --(9.030,7.178)--(9.097,7.236)--(9.164,7.295)--(9.231,7.353)--(9.299,7.412)--(9.366,7.470)%
  --(9.433,7.529)--(9.500,7.587)--(9.568,7.646)--(9.635,7.704)--(9.702,7.763)--(9.769,7.821)%
  --(9.837,7.880)--(9.904,7.938)--(9.971,7.997)--(10.038,8.055)--(10.106,8.114);
\gpcolor{gp lt color 3}
\gpsetpointsize{4.00}
\gppoint{gp mark 3}{(9.500,7.647)}
\gppoint{gp mark 3}{(9.769,7.251)}
\gppoint{gp mark 3}{(8.626,6.901)}
\gppoint{gp mark 3}{(8.357,6.530)}
\gppoint{gp mark 3}{(7.954,6.074)}
\gppoint{gp mark 3}{(7.348,5.792)}
\gppoint{gp mark 3}{(7.214,5.416)}
\gppoint{gp mark 3}{(6.609,4.924)}
\gppoint{gp mark 3}{(6.004,4.556)}
\gppoint{gp mark 3}{(5.600,4.185)}
\gppoint{gp mark 3}{(5.129,3.810)}
\gppoint{gp mark 3}{(4.726,3.443)}
\gppoint{gp mark 3}{(4.322,3.089)}
\gppoint{gp mark 3}{(3.986,2.696)}
\gppoint{gp mark 3}{(3.448,2.327)}
\gpcolor{gp lt color 3}
\gppoint{gp mark 4}{(9.635,1.604)}
\gppoint{gp mark 4}{(9.097,1.579)}
\gppoint{gp mark 4}{(8.626,1.558)}
\gppoint{gp mark 4}{(10.106,1.531)}
\gppoint{gp mark 4}{(7.752,1.509)}
\gppoint{gp mark 4}{(7.416,1.488)}
\gppoint{gp mark 4}{(6.945,1.466)}
\gppoint{gp mark 4}{(6.474,1.444)}
\gppoint{gp mark 4}{(6.071,1.424)}
\gppoint{gp mark 4}{(7.281,1.408)}
\gppoint{gp mark 4}{(5.331,1.383)}
\gppoint{gp mark 4}{(4.726,1.358)}
\gppoint{gp mark 4}{(4.322,1.337)}
\gppoint{gp mark 4}{(3.919,1.315)}
\gppoint{gp mark 4}{(3.515,1.293)}
\gpcolor{gp lt color border}
\draw[gp path] (2.843,8.381)--(2.843,1.229)--(10.240,1.229);
\gpdefrectangularnode{gp plot 1}{\pgfpoint{2.843cm}{1.229cm}}{\pgfpoint{10.240cm}{8.381cm}}
\end{tikzpicture}
  }
 \vspace{-.55cm}
 \caption{\smaller{Overhead cost}}
  \label{fig:hvrrunningtime}
 \end{subfigure}
 \caption{Refinement checker for the Hypervisor}
 \end{figure}

\vspace{-.2cm}
\paragraph{Mutations:} We manually injected mutations in the
virtualized system to evaluate the effectiveness of refinement testing
to detect bugs. In this case study, we restrict our attentions to
mutations in the hypervisor component of the virtualized system. The
mutations are informally classified into four classes: (1)
\emph{Save/Restore:} mutations injected in save and restore routines
of the host processor state in the hypervisor; (2) \emph{Emulation of
  a privilege instruction:} mutations injected in hypervisor routines
that emulate the execution of privileged instructions; (3)
\emph{Shadow Page Table:} mutations injected in hypervisor routines
that sync the shadow page-table with the guest page-table; (4)
\emph{Address Calculation:} mutations injected in the address
calculations performed during the translation of guest virtual address
to host physical address. Table~\ref{tab:hyp-mutations} shows that
refinement testing found all of these errors.

 \vspace{-.7cm}

In Figure~\ref{fig:hvrrunningtime}, we plot the running time for
simulating the virtualized system with ($y$-axis) and without
refinement checker ($x$-axis), where the number of steps ranges from
10,000 to 100,000. The slope of the fitting line (blue) is $\sim
65$. The reason for the large slow down is the sub-procedure in the
refinement map used to compute the guest memory from the host memory;
it traverses the host memory and extracts the guest memory for each
guest in the system. If the size of the memory is large, which is
often the case, this is prohibitively expensive. To reduce the cost of
computing the refinement map, we add a history variable to the
virtualized system that records the guests accesses to the host
memory. Note that augmenting the system with the history variable does
not modify its observable behavior. We then modify the refinement map
to use the history variable to construct the updated guest memory from
the initial guest memory. We again compare the running times for
simulating the virtualized system with and without the modified
refinement map. In this case the slope of the fitting line (green) is
$\sim 3.6$, \ie we get over 18 times speed up in the running time of
the virtualized system with the modified WEB refinement check. This
experiment reaffirms that the refinement map plays a crucial role
designing an efficient refinement checker.

Notice that in a property-based testing methodology, we would have to
describe a property for each of the mutations described in
Table~\ref{tab:hyp-mutations}. Furthermore, even such a list of
properties is not complete, \eg it does not check that after handling
an exception, the hypervisor eventually resumes executing the guest on
the host processor directly. In contrast, WEB refinement check
accounts for such liveness properties of the hypervisor.


\subsection{RISC-V Based Pipelined Processors}
\label{sec:riscv}

We next evaluate the effectiveness of our methodology to check
functional correctness of pipeline processors based on RISC-V
instruction set specfication~\cite{riscv}. RISC-V is an open source
specification, which is being widely adopted both for educational and
commercial use. Among several open source implementations available,
we choose to evaluate our methodology using the Sodor processor
collection, a collection of 2, 3, and 5-stage pipeline processor
models~\cite{riscvsodor}. These processors implement the RISC-V 32-bit
integer base user-level instructions using different
micro-architectural features. The models are described in Chisel, a
hardware description language embedded in Scala; it then can be
translated to either a fast bit and cycle-accurate \emph{C++}
simulator or a low-level Verilog description suitable for hardware
emulation~\cite{bachrach2012chisel}. The open source community has
also developed Spike, an executable model of the RISC-V instruction
set specification, that serves as ``golden standard'' for
execution. In addition, the community has also developed an extensive
testing infrastructure consisting of several directed tests and
benchmark programs. These tests are described in \emph{C} and can be
compiled to execute on both the Spike and Sodor processors. We
evaluate the applicability of the refinement-based testing methodology
to check functional correctness of the 5-stage pipeline Sodor
processor (5SP). It is a single-issue in-order pipeline processor
that supports full bypassing between functional units. We use Spike as
a high-level abstract specification for checking functional
correctness of the Sodor processor.

As in the previous two case studies (Section \ref{sec:pipeproc} and
\ref{sec:hypervisor}), the correctness of the 5SP processor can be
specified by a single WEB refinement check. Moreover, the 5SP
processor and the Spike are deterministic systems and the later does
not stutter; hence we can use the simplified WEB refinement conjecture
(Figure~\ref{fig:detwebrefinement}) to design an efficient runtime
checker.

Before we discuss the challenges in implementation of the refinement
checker for the 5SP processor, we first define the refinement map $r$,
and the ranking function $\rankt$. Refinement map $r$ takes as an
input a state of the 5SP processor, invalidates all partially executed
instructions, then based on the number of partially executed
instructions it moves the program counter back, and finally projects
all programmer-observable components to return a state of the
Spike. In order to efficiently compute the refinement map, we
instrument the processor implementation using history variables. We
add a \emph{valid bit} corresponding to each pipeline latch; the valid
bit is set to true if the instruction in the pipeline stage is
valid. In addition, we also record the program counter for each
partially executed instruction in the pipeline. The instrumentation of
the 5SP processor to add these history variables required only a few
additional lines of code. Next, we define the ranking function
$\rankt$. Given a state of the 5SP processor, its rank is defined by
the minimum number of steps the processor needs to commit a partially
executed instruction, \ie take an observable step.

Now we are ready to discuss the implementation of the WEB refinement
checker. Notice that the use of Algorithm~\ref{alg:general} to
implement the refinement checker requires a mechanism to control the
stepping of both the 5SP processor and Spike (Line 3 and 4 in
Algorithm~\ref{alg:general}). Though possible (as in the previous two
case studies), it would involve substantial modifications to the
existing simulation and testing infrastructure of Sodor processors, an
undesirable scenario for adoption of the refinement-based testing
methodology. Therefore, we design an alternative implementation of the
refinement checker that decouples the execution of the 5SP processor,
the Spike simulator, and the computation of WEB refinement check. We
spawn three independent threads: (1) the concrete system (Sodor
processor) simulation; (2) the abstract system (Spike) simulation; and
(3) the refinement checker. Thread (1) and (2) acts as producers and
thread (3) acts as the consumer; the communication between the
producers and the consumer is asynchronous. This alternative
implementation only requires a mechanism in the simulator to
communicate its state to the refinement checker in each cycle. For the
5SP simulator and the Spike simulator that are implemented in C++,
this is done by simply adding \emph{printf} statements.

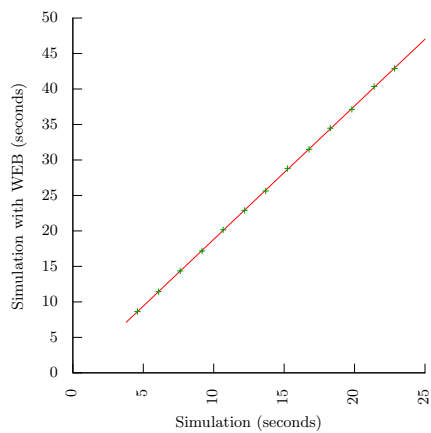
\begin{figure}[h]
\vspace{-.3cm}
  \centering
  \captionsetup{font=footnotesize}
  \scalebox{.66}{
    \begin{tikzpicture}[gnuplot]
\gpcolor{gp lt color border}
\gpsetlinetype{gp lt border}
\gpsetlinewidth{1.00}
\draw[gp path] (3.087,1.229)--(3.267,1.229);
\node[gp node right] at (2.903,1.229) { 0};
\draw[gp path] (3.087,1.944)--(3.267,1.944);
\node[gp node right] at (2.903,1.944) { 5};
\draw[gp path] (3.087,2.659)--(3.267,2.659);
\node[gp node right] at (2.903,2.659) { 10};
\draw[gp path] (3.087,3.375)--(3.267,3.375);
\node[gp node right] at (2.903,3.375) { 15};
\draw[gp path] (3.087,4.090)--(3.267,4.090);
\node[gp node right] at (2.903,4.090) { 20};
\draw[gp path] (3.087,4.805)--(3.267,4.805);
\node[gp node right] at (2.903,4.805) { 25};
\draw[gp path] (3.087,5.520)--(3.267,5.520);
\node[gp node right] at (2.903,5.520) { 30};
\draw[gp path] (3.087,6.235)--(3.267,6.235);
\node[gp node right] at (2.903,6.235) { 35};
\draw[gp path] (3.087,6.951)--(3.267,6.951);
\node[gp node right] at (2.903,6.951) { 40};
\draw[gp path] (3.087,7.666)--(3.267,7.666);
\node[gp node right] at (2.903,7.666) { 45};
\draw[gp path] (3.087,8.381)--(3.267,8.381);
\node[gp node right] at (2.903,8.381) { 50};
\draw[gp path] (3.087,1.229)--(3.087,1.409);
\node[gp node right,rotate=-270] at (3.087,1.045) { 0};
\draw[gp path] (4.506,1.229)--(4.506,1.409);
\node[gp node right,rotate=-270] at (4.506,1.045) { 5};
\draw[gp path] (5.924,1.229)--(5.924,1.409);
\node[gp node right,rotate=-270] at (5.924,1.045) { 10};
\draw[gp path] (7.343,1.229)--(7.343,1.409);
\node[gp node right,rotate=-270] at (7.343,1.045) { 15};
\draw[gp path] (8.761,1.229)--(8.761,1.409);
\node[gp node right,rotate=-270] at (8.761,1.045) { 20};
\draw[gp path] (10.180,1.229)--(10.180,1.409);
\node[gp node right,rotate=-270] at (10.180,1.045) { 25};
\draw[gp path] (3.087,8.381)--(3.087,1.229)--(10.180,1.229);
\node[gp node center,rotate=-270] at (2.013,4.805) {Simulation with WEB (seconds)};
\node[gp node center] at (6.633,0.215) {Simulation (seconds)};
\gpcolor{gp lt color 0}
\gpsetlinetype{gp lt plot 0}
\draw[gp path] (4.162,2.249)--(4.233,2.316)%
  --(4.305,2.384)--(4.377,2.452)--(4.448,2.520)--(4.520,2.588)--(4.592,2.656)--(4.663,2.724)%
  --(4.735,2.792)--(4.807,2.860)--(4.878,2.928)--(4.950,2.996)--(5.021,3.064)--(5.093,3.132)%
  --(5.165,3.200)--(5.236,3.268)--(5.308,3.335)--(5.380,3.403)--(5.451,3.471)--(5.523,3.539)%
  --(5.595,3.607)--(5.666,3.675)--(5.738,3.743)--(5.810,3.811)--(5.881,3.879)--(5.953,3.947)%
  --(6.025,4.015)--(6.096,4.083)--(6.168,4.151)--(6.239,4.219)--(6.311,4.287)--(6.383,4.354)%
  --(6.454,4.422)--(6.526,4.490)--(6.598,4.558)--(6.669,4.626)--(6.741,4.694)--(6.813,4.762)%
  --(6.884,4.830)--(6.956,4.898)--(7.028,4.966)--(7.099,5.034)--(7.171,5.102)--(7.242,5.170)%
  --(7.314,5.238)--(7.386,5.306)--(7.457,5.373)--(7.529,5.441)--(7.601,5.509)--(7.672,5.577)%
  --(7.744,5.645)--(7.816,5.713)--(7.887,5.781)--(7.959,5.849)--(8.031,5.917)--(8.102,5.985)%
  --(8.174,6.053)--(8.246,6.121)--(8.317,6.189)--(8.389,6.257)--(8.460,6.325)--(8.532,6.392)%
  --(8.604,6.460)--(8.675,6.528)--(8.747,6.596)--(8.819,6.664)--(8.890,6.732)--(8.962,6.800)%
  --(9.034,6.868)--(9.105,6.936)--(9.177,7.004)--(9.249,7.072)--(9.320,7.140)--(9.392,7.208)%
  --(9.464,7.276)--(9.535,7.344)--(9.607,7.411)--(9.678,7.479)--(9.750,7.547)--(9.822,7.615)%
  --(9.893,7.683)--(9.965,7.751)--(10.037,7.819)--(10.108,7.887)--(10.180,7.955);
\gpcolor{gp lt color 1}
\gpsetpointsize{4.00}
\gppoint{gp mark 1}{(4.389,2.466)}
\gppoint{gp mark 1}{(4.815,2.865)}
\gppoint{gp mark 1}{(5.249,3.283)}
\gppoint{gp mark 1}{(5.689,3.686)}
\gppoint{gp mark 1}{(6.114,4.113)}
\gppoint{gp mark 1}{(6.546,4.505)}
\gppoint{gp mark 1}{(6.971,4.895)}
\gppoint{gp mark 1}{(7.411,5.347)}
\gppoint{gp mark 1}{(7.845,5.735)}
\gppoint{gp mark 1}{(8.273,6.161)}
\gppoint{gp mark 1}{(8.707,6.540)}
\gppoint{gp mark 1}{(9.156,7.002)}
\gppoint{gp mark 1}{(9.567,7.364)}
\gpcolor{gp lt color border}
\gpsetlinetype{gp lt border}
\draw[gp path] (3.087,8.381)--(3.087,1.229)--(10.180,1.229);
\gpdefrectangularnode{gp plot 1}{\pgfpoint{3.087cm}{1.229cm}}{\pgfpoint{10.180cm}{8.381cm}}
\end{tikzpicture}
  }
  \caption{\smaller{Overhead cost for the 5-stage pipeline Sodor
      processor}}
  \label{fig:sodorrunningtime}
\end{figure}

We implement the WEB refinement checker in Python. The refinement
checker spawns off the 5SP simulator and the Spike simulator as
independent processes. {In each cycle, the simulator write the state
  of the system into a FIFO.} The refinement checker reads the state
of the system from the FIFOs and checks if the WEB refinement
conjecture holds. Recall that the WEB refinement check is local and
thus only needs to analyze the current state of the systems and their
successors. In case the checker detects a violation, it outputs the
offending state and quits. Else it evicts the analyzed states of the
system from the respective FIFOs. In
Figure~\ref{fig:sodorrunningtime}, we plot the running time for
simulating the 5SP with ($y$-axis) and without ($x$-axis) the
refinement checker; the number of cycles range from 300000 to
1500000. The slope of the fitting line is $\sim$1.8, which amounts to
a slowdown of approximately 80\%. The memory footprint of the
refinement checker is small; this is because the WEB refinement check
is local and only requires us to analyze states and their
successors. Also notice that points on the plot are either on the
fitting line, or near it; this indicates that the overhead cost of the
refinement checker does not increase with increase in the number of
cycles in a simulation run. This is desirable property in long running
simulations.

To evaluate the effectiveness of the refinement-based testing
methodology to detect bugs, we introduced 20 mutations in 5-state
pipeline Sodor processor. These mutations are similar to the ones
introduced in the MA machine (Table~\ref{tab:mamutations}). We use the
generic programs like quick sort and the tower of Hanoi puzzle in the
RISC-V test suite~\cite{riscvtests}. With this test suite, the
refinement checker could detect all mutations that could result in
functional violations.

\vspace{-.3cm}
\section{Discussion and Related Work}
\label{sec:discussion}

In this section, we discuss how our refinement-based testing overcomes
several major drawbacks of the standard testing methods.  In the
previous section, we showed that the formal refinement conjecture can
be compiled into an efficient runtime check that can be used to detect
bugs in a dynamic validation workflow. Furthermore, unlike temporal
properties, that can span multiple steps of the machine,
WEB-refinement check is local, \ie we only check the relation between
a state and its successor. This allows us to limit the overhead cost
without compromising on the expressivity. The overhead of a refinement
checker during simulation can further be improved by exploiting domain
specific knowledge (deterministic machines), augmenting the machine
with appropriate history variables, choosing an efficient refinement
map, and checking only for the safety or the liveness component of the
refinement check. In contrast, in a property-based methodology,
properties are expressed in an assertion language, like SVA and PSL,
and compiled into a (finite-state) automaton. In the worst case, the
automaton can be exponential in the size of the property and the
overhead of checking the property during simulation can be
prohibitively expensive. In~\cite{turumella2008assertion}, it is
reported that depending on the number and the kind of properties they
express, property checking can degrade simulation speed anywhere from
25\% to 100\%.

A major concern with current testing methods is how to decide if the
set of properties or tests is ``complete.''  An alert reader would
have noticed that the set of properties corresponding to the mutations
in Section~\ref{sec:pipeproc} does not include the one that checks
that if the pipeline in the MA machine stalls, say due to data
dependency, it is eventually un-stalled and resumes fetching new
instructions. With this incomplete set of properties, an MA machine
that never resumes fetching new instructions will be incorrectly
declared as functionally correct. As systems grow in complexity, it
becomes increasingly difficult to answer the question of
completeness. In~\cite{kaivola2009replacing}, the authors describe
their experience in a twenty person year effort to formally verify the
Execution Cluster, a component of the Intel Core i7 microprocessor
that is responsible for functional behavior of all
microinstructions. It is reported that at the end of their effort they
only missed detecting five bugs found in the fabricated
processor. They attributed three of these misses to ``incorrect formal
specification'' and two of these misses to ``formal verification work
not being completed early enough''. Both reasons are indicative of the
difficulty in specifying a correct and complete set of properties for
complex designs. An industry-wide survey in 2014 shows that over 50\%
of respins are due to logic or functional errors; 40\% of these
functional flaws are attributed to incorrect/incomplete
specifications~\cite{foster2015}. In contrast, WEB-refinement
completely characterizes the functional correctness of the design and
therefore does not suffer from the incompleteness problem.

In section~\ref{sec:testing}, the locality of WEB refinement
conjecture (reasoning only about states and their successors) played a
key part in designing an efficient runtime checker. This quality of
WEB refinement also makes it amenable for automated reasoning using
existing formal verification
tools~\cite{manolios2004automatic,manolios2005complete}. Hence,
refinement-based approach satisfies a key requirement: a specification
must be effectively analyzed both in a dynamic validation workflow as
well as in a formal verification workflow. The advantage of such a
specification is that it acts as a bridge between dynamic validation
and formal verification.

Finally, we note that our refinement-based testing can be used in a
variety of ways to combine dynamic and static verification
methods. For example, we might prove that a high-level, abstract
system satisfies a set of desired properties. This is much easier to
do at the abstract level than at the concrete level.  If we can
statically show that concrete system refines the high-level system,
say using WEB refinement, then, since WEB refinement preserves all the
CTL$^{*}$\textbackslash X properties (both safety and liveness
properties), we can infer that the concrete system satisfies all the
temporal properties that were validated for the abstract system. Such
a proof may be difficult to perform. Instead, we can use
refinement-based testing to gather evidence that the concrete system
refines the abstract system.  .

\vspace{-.3cm}
\section{Conclusions and Future Work}
\label{sec:conc}
In this paper we introduced a refinement-based methodology for testing
functional correctness of hardware systems and low-level software. We
showed that our methodology is effective in detecting bugs and can be
easily integrated in existing simulation workflows.  We introduced an
effective algorithm for checking WEB refinement during simulation. For
future work, we plan to explore the use of refinement-based testing
to check the functional correctness of nondeterministic systems.
Also the notion of WEB refinement is compositional and supports
top-down stepwise refinement methodology; we plan to explore how to
fully exploit the compositionality of refinement to test complex,
hierarchical designs with external IP components.

\vspace{-.15cm}
\bibliographystyle{abbrv} 
\bibliography{paper-reduced}
\end{document}